\newcommand{\un}{~\mathrm}

\documentclass[aps,pre,showpacs,twocolumn]{revtex4}
\usepackage{amsmath,epsfig,amsfonts,graphicx,color,ulem}
\usepackage{graphicx}

\begin{document}

\title{Nonlinear Waves in Disordered Diatomic Granular Chains}
\author{
Laurent Ponson$^{1}$,
Nicholas Boechler$^{1}$,
Yi Ming Lai$^{2}$,
Mason A. Porter$^{3}$,
P.~G. Kevrekidis$^{4}$, and
Chiara Daraio$^{1}$
}

\affiliation{
$^1$Graduate Aerospace Laboratories (GALCIT), California Institute of Technology, Pasadena, CA 91125, USA \\
$^2$Oxford Centre for Collaborative Applied Mathematics, Mathematical Institute, University of Oxford, OX1 3LB, UK \\
$^3$Oxford Centre for Industrial and Applied Mathematics, Mathematical Institute, University of Oxford, OX1 3LB, UK \\
$^4$Department of Mathematics and Statistics, University of Massachusetts, Amherst MA 01003-4515, USA 
}


\begin{abstract}

We investigate the propagation and scattering of highly nonlinear waves in disordered granular chains composed of diatomic (two-mass) units of spheres that interact via Hertzian contact.  Using ideas from statistical mechanics, we consider each diatomic unit to be a \textquotedblleft spin", so that a granular chain can be viewed as a spin chain composed of units that are each oriented in one of two possible ways.  Experiments and numerical simulations both reveal the existence of two different mechanisms of wave propagation: In low-disorder chains, we observe the propagation of a solitary pulse with exponentially decaying amplitude.  Beyond a critical level of disorder, the wave amplitude instead decays as a power law, and the wave transmission becomes insensitive to the level of disorder.  
We characterize the spatio-temporal structure of the wave in both propagation regimes and propose a simple theoretical interpretation for such a transition.  Our investigation suggests that an elastic spin chain can be used as a model system to investigate the role of heterogeneities in the propagation of highly nonlinear waves.


\end{abstract}


\maketitle


\section{Introduction} \label{sec1}


A recent focal point of studies on nonlinear oscillators is one-dimensional (1D) granular crystals, which consist of closely packed chains of elastically colliding particles.  Because of the contact interaction between particles, granular crystals are characterized by a highly nonlinear dynamic response. This has inspired numerous studies of the interplay between nonlinearity and discreteness \cite{nesterenko1,sen08}.  Granular crystals can be created from numerous material types and sizes, making their properties extremely tunable \cite{nesterenko1,sen08,nesterenko2,coste97}.  Because they provide an experimental setting that is both tractable and flexible, granular crystals are an excellent testbed for investigating the effects of structural and material heterogeneities on nonlinear wave dynamics. Recent studies have examined the role of defects \cite{hascoet}, interfaces between two different types of particles \cite{dar05b,dar06}, decorated and/or tapered chains \cite{doney06,harbola09}, chains of diatomic and triatomic units \cite{dimer,herbdimer,molinari09}, and quasiperiodic and random configurations \cite{senrandom07,chen07,fernando}. The tunability of granular crystals is valuable not only for basic studies of the underlying physics but also in potential engineering applications, including shock and energy absorbing layers \cite{dar06,hong05,fernando}, sound focusing devices and delay lines \cite{spandoni09}, actuators \cite{dev08}, vibration absorption layers \cite{herbdimer}, and sound scramblers \cite{dar05,dar05b}. 

An important context for the present paper is the consideration of disordered settings in nonlinear oscillator chains.  Some of the most prominent recent investigations in this area have generalized to weakly nonlinear settings \cite{andersonstuff} the ideas of P.~W. Anderson, who showed theoretically that the diffusion of waves is curtailed in linear media that contain sufficient randomness induced by defects or impurities \cite{anderson58,andersonreview}.  Research on the Anderson transition is extremely exciting and is germane to the investigation of disordered systems more generally, but our goal in this study departs somewhat from this thread, as we seek to investigate order-disorder transitions in \textit{strongly nonlinear} media such as granular crystals. This is a key challenge in the study of nonlinear chains \cite{kuzflach}.

In this paper, we investigate 1D granular crystals that consist of chains of elastic spheres that interact via Hertzian contacts.  The chains are composed of  diatomic units: two-mass cells (henceforth called \textquotedblleft dimers") that each consist of one particle of one type (\textquotedblleft type 1") and one particle with different properties (\textquotedblleft type 2").  Each dimer can be arranged in one of two orientations: type 1 followed by type 2 or vice versa.  To quantify the nature of the disorder and define the chain configuration, we borrow an idea from statistical physics and treat each cell as a \textquotedblleft spin" (where, for example, spin \textquotedblleft up" is characterized by a dimer composed of a type 1 particle followed by a type 2 particle, and spin \textquotedblleft down" is characterized by a dimer composed of a type 2 particle followed by a type 1 particle).  This allows a novel perspective on investigations of granular crystals, as one can examine the wave propagation dynamics as a function of an order parameter, which we define using the ratio of the number of spins oriented in the same direction to the total number of spins.  Borrowing another technique of statistical physics (which, in particular, is used in the context of disordered systems), we take an ensemble average of quantities that characterize the wave propagation (amplitude, wave velocity, etc.) over chain configurations with the same level of disorder.  Accordingly, as disorder increases, we are able to show that the wave transmission decreases rapidly, saturating to a value that is independent of the chain heterogeneity.

To characterize the order-disorder transition, we analyze local properties of the wave (namely, its amplitude and spatial structure) during its propagation.  This reveals two propagation mechanisms: We find solitary waves below a threshold level of disorder and delocalized waves above it.  We use the term \textquotedblleft wave delocalization" to mean that the spatial structure of the solitary waves---which extend over only a few beads and are observed in the low-disorder regime---disappears with increasing disorder, giving rise to a significantly more extended wave structure that is present over a long portion of the chain.  We also investigate the effects of chain length and material properties
and propose a simple theoretical description---based on a scattering mechanism across individual defects in the chain---that accounts for most of the wave properties.

The remainder of this paper is organized as follows.  We detail our definition of elastic spin chains in Section \ref{sec2} and discuss the setup for our experiments and numerical computations, respectively, in Sections \ref{sec3} and \ref{sec4}.  We then discuss in more detail the order-disorder transition and the dynamics in the low-disorder and high-disorder regimes in Sections \ref{sec5} and \ref{sec6}.  We discuss our results in Section \ref{Section_discussion} and conclude in Section \ref{sec7}.



\section{Elastic Spins and Disorder} \label{sec2}

The concept of spin has been used successfully to describe myriad physical phenomena, such as magnetization and glassiness \cite{sethnabook}.  In this paper, we apply this idea to analyze order versus disorder in granular crystals.  Fully ordered diatomic chains allow the propagation of solitary waves \cite{dimer}. However, an orientation reversal of even one cell causes a defect in the system, leading to complicated dynamics such as partial wave reflections and radiation shedding \cite{sen98,fernando}. Increasing the heterogeneity of the chain further via random arrangements of cells alters the dynamics of wave propagation drastically and necessitates a different approach.
To understand such heterogeneous granular chains, we characterize the level of disorder in the chain by using the diagnostic
\begin{equation}
	D = 1 - \frac{|N_{\mbox{up}} - N_{\mbox{down}}|}{N_{\mbox{up}} + N_{\mbox{down}}}\,, \label{mag}
\end{equation}
where $N_{\mbox{up}}$ is the number of dimers (spins) with one orientation (called \textquotedblleft up") and $N_{\mbox{down}}$ is the number of dimers with the opposite spin (called \textquotedblleft down").  
The total number of spins, $N = N_{\mbox{up}} + N_{\mbox{down}}$, is equal to one half of the number of particles in the chain. By convention, we say that a dimer has spin ``up'' if the heavier particle is nearer the direction of the original incident wave. The definition (\ref{mag}) guarantees that maximum order---the case studied in Ref.~\cite{dimer}, in which all dimers have the same orientation---occurs when $D = 0$.  Minimum order (and hence maximum disorder) occurs when $N_{\mbox{up}}=N_{\mbox{down}}$, which yields $D = 1$ \footnote{There are configurations (e.g., a periodic sequence of dimers with alternating spin orientations) for which $D = 1$ gives maximum order because of higher-order correlations. Upon averaging, these contribute little to the expected dynamics due to their low probability of occurrence.}. The quantity $D$ thereby measures the amount of heterogeneity in the granular chain.  It is worth remarking that $D = 2N_{{\rm defect}}/N$, where $N_{{\rm defect}}$ is the number of ``defects" (situations in which consecutive spins have opposite orientation) in the chain.
 
One obtains a given value of $D$ for many different possible combinations of dimers, so $D$ indicates only the presence of \textquotedblleft defective" spins and not their location. To account for this, we average over multiple disordered configurations with equal values of $D$. The effect of the level of disorder $D$ on the transmission of nonlinear waves through the chain is the central point of this study. As $D$ increases, we expect the amplitude of the transmitted force to decrease. However, our study reveals (see the discussion below) that this is true only for sufficiently small values of the parameter $D$, for which the solitary-like wave structure is preserved.  In contrast, we observe waves with delocalized structure at larger values of $D$, and the wave propagation is no longer affected by the level of heterogeneity in the chain.  The issues that we address below are (1) when this transition occurs and (2) how we can characterize and interpret both of these regimes.


\section{Experimental Setup} \label{sec3}

\begin{figure}[tbp]
\begin{center}
\vspace{-1.0cm}
\includegraphics[width = 8.5cm]{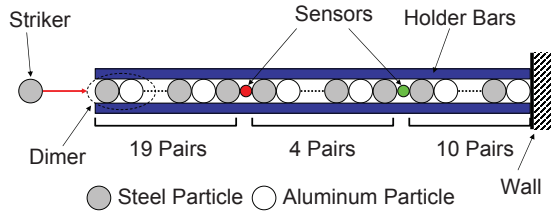}
\end{center}
\vspace{-2.0cm}
\caption{[Color online] Schematic of the experimental setup.}
\label{setup}
\end{figure}

To experimentally investigate the order-disorder transition in granular chains, we construct chains of dimer cells composed of units of spherical beads of two different materials and/or sizes. For the experimental results in our main discussion, we use 4.76 mm radius stainless steel (non-magnetic, 316 type) and aluminum (Al; 2017-T4 type) spheres. The steel and Al beads, respectively, have masses 3.63 g and 1.26 g, elastic moduli $193 \un{GPa}$ and $72.4 \un{GPa}$, and Poisson ratios 0.30 and 0.33 \cite{316,alum}. We examine chains with various sequences of up and down cells and assemble each chain in a horizontal setup [see Fig.~\ref{setup}(a)] that is composed of two steel bars clamped on a sine plate. We ensure contact (and hence 1D impact and wave propagation) between the particles by tilting the guide slightly (3.5 degrees). The chain is 31 cells long.

We generate the incident solitary waves by impacting the chain with a steel particle \textquotedblleft striker" (of the same size as the steel particles in the chain), which is launched down a ramp.  Based on the drop height of the striker, we calculate its impact velocity to be about $0.5$ m/s for these experiments. To visualize the traveling waves directly, we place piezo sensors ($RC \sim 10^3 \,\mu\mbox{s}$, Piezo Systems Inc.) inside small (2.38 mm radius) stainless steel beads (non-magnetic, 316 type) \cite{316,ricardo09,dimer} that we use in place of the light Al particle in the 20th and 25th cells. The sensors are kept in place by thin support plates with circular holes whose diameter is slightly larger than that of the beads. 

Using this setup, we consider values of the parameter $D$, which characterizes the level of disorder in the first 20 pairs of the chain (see the discussion above), ranging from 0 to 1 in increments of 0.1. For each value, we consider 3 different dimer cell arrangements, and we average the results from 3 striker drops for each such arrangement.  We also performed similar experiments with other types of diatomic chains---including one composed of large stainless steel (radius 4.76 mm) and small (2.38 mm) stainless steel spheres and another composed of small stainless steel and small PTFE (also 2.38 mm) spheres.  We obtained consistent results with all of the considered configurations.  In our discussion below, we present the steel-aluminum example.



\section{Numerical Setup} \label{sec4}

We model a chain of $2N$ spherical beads as a 1D lattice with (conservative) Hertzian interactions between particles \cite{nesterenko1,sen08}:
\begin{align}
	\ddot{y}_j &= \frac{A_{j-1,j}}{m_j}\delta_{j}^{3/2} - \frac{A_{j,j+1}}{m_j}\delta_{j+1}^{3/2}\,, \notag \\
A_{j,j+1} &= \frac{4E_{j}E_{j+1}\left(\frac{R_jR_{j+1}}{R_j + R_{j+1}}\right)^{1/2}}{3\left[E_{j+1}\left(1-\nu_{j}^2\right) + E_j\left(1-\nu_{j+1}^2\right)\right]}\,, \label{motion}
\end{align}
where $j \in \{1,\cdots,2N\}$, $y_j$ is the coordinate of the center of the $j$th particle measured from its equilibrium position, $\delta_j \equiv \mbox{max}\{y_{j-1} - y_{j},0\}$ for $j \in \{2, 3, \dots, 2N\}$, $\delta_1 \equiv 0$, $\delta_{2N+1} \equiv \mbox{max}\{y_{2N},0\}$, $E_j$ is the elastic modulus of the $j$th particle, $\nu_j$ is its Poisson ratio, $m_j$ is its mass, and $R_j$ is its radius. The particle $j = 0$ represents the striker, and the $(2N+1)$th particle represents the wall. The initial velocity of the striker is determined from experiment, and all other particles start at rest in their equilibrium positions.

\section{Comparison between Experiments and Numerics} \label{sec5}

\begin{figure}[tbp]
\centering{
(a) \includegraphics[width = 9.0 cm]{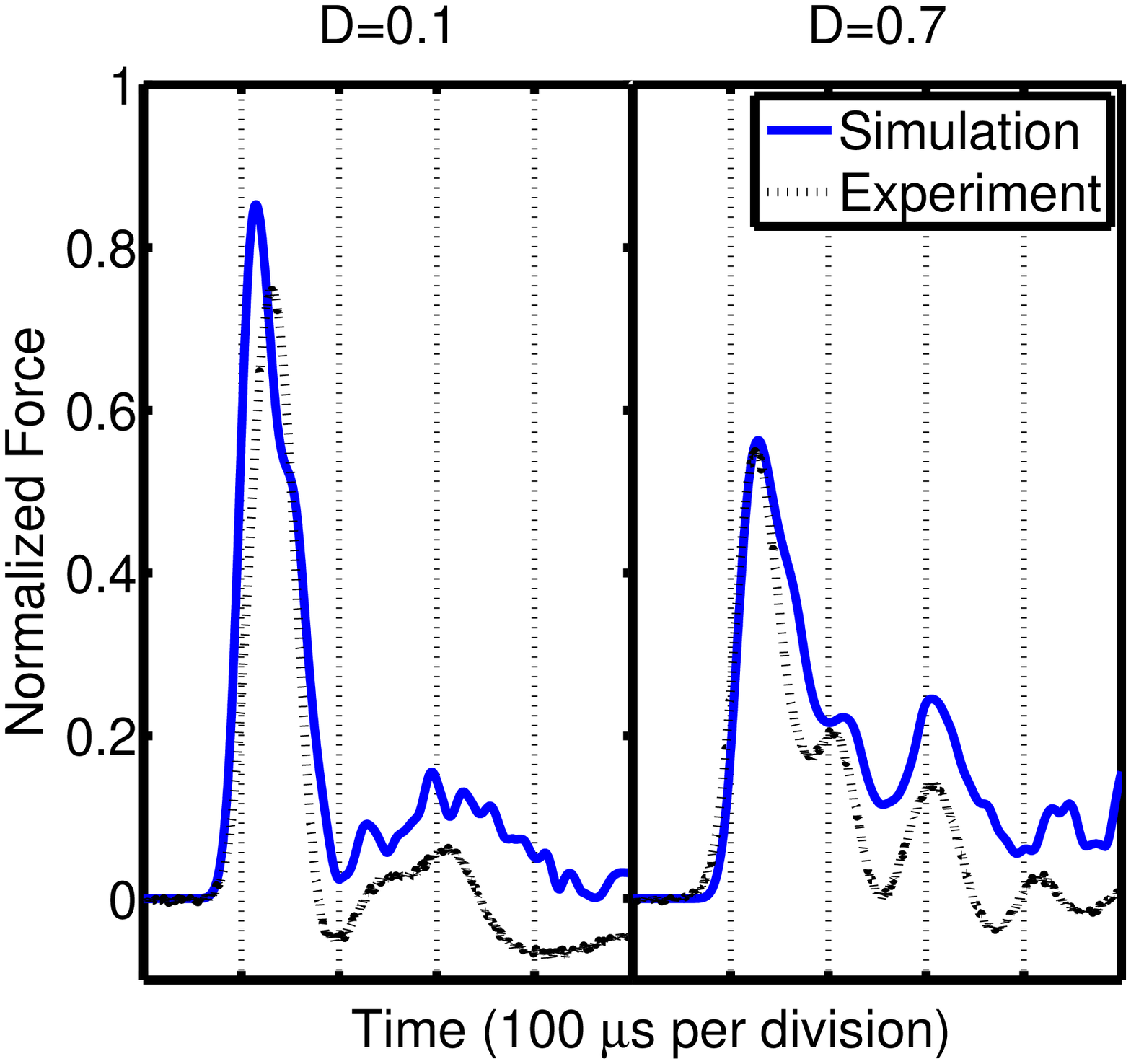}
(b) \includegraphics[width = 9.0 cm]{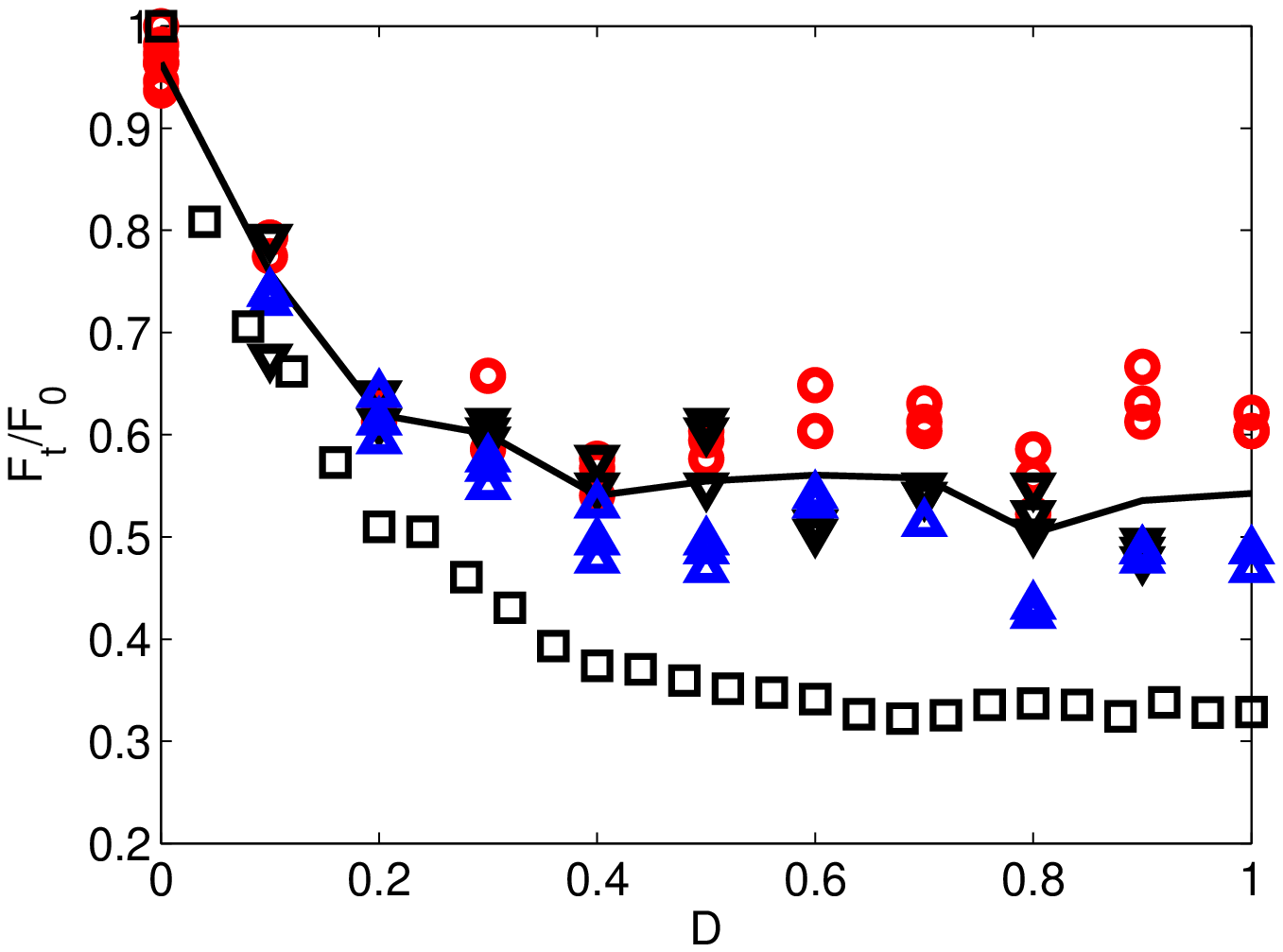}
}
\caption{[Color online] (a) Averaged experimental and numerical plots of force versus time (as measured by a sensor in the 20th cell of a steel:Al chain) for configurations with a low ($D = 0.1$) and a high ($D = 0.7$) level of disorder.  We normalize the force based on the peak value for the perfectly ordered system (for which $D = 0$). (b) Normalized peak transmitted force (sensor in 20th cell; steel:Al chain) as a function of disorder. Each color/shape represents a different configuration of equivalent disorder, and each marker corresponds to a separate experimental run. The solid curve gives the median transmitted force value for each level of disorder, and the squares (showing the same qualitative behavior) give the results of the numerical runs that correspond to the median experimental configurations.
}
\label{Fig2}
\end{figure}

We compare numerical simulations of Eq.~(\ref{motion}) with experiments using the same type and quantity of particles composing the chain, the same ``spin configuration'', and the same excitation input. Once the striker hits the first bead, a compression wave propagates through the chain.  We measure the force of the propagating wave using a sensor that we placed in the 20th cell. In Fig.~\ref{Fig2}(a), we show plots of force versus time for both low and high values of $D$ for the experiments and the numerical computations. Panel (b) of Fig.~\ref{Fig2} shows the peak amplitude of the transmitted wave as a function of $D$ for both experiments and numerical simulations. In both plots, we show the normalized force (which is given by the measured force divided by the peak force that we obtained in the perfectly ordered case). We consider three arrangements corresponding to each $D$ value, and we acquire three independent measurements for each arrangement to verify experimental reproducibility. In Fig.~\ref{Fig2}(a), we show the force-time history averaged over all three different dimer cell arrangements at each level of disorder. In Fig.~\ref{Fig2}(b), we show the peak normalized transmitted force.  Each color/shape corresponds to a given spin configuration at the stated value of $D$, and each marker represents a single experimental run. The solid curve gives the median transmitted force value for each level of disorder.  The squares give the results of the numerical runs that correspond to the median experimental configurations. 

We observe good qualitative agreement between numerics and experiments.  Note, in particular, the decrease of the peak force as $D$ becomes larger. Because the numerics and experiments agree qualitatively---the small quantitative difference arises from experimental dissipation that is not modeled in Eq.~(\ref{motion})---we focus the remainder of our discussion on the former.

\section{Order-Disorder Transition: From Solitary to Delocalized Waves} \label{sec6}

\subsection{Wave Transmission Model}

\begin{figure}[tbp]
\centering{
\includegraphics[width = 8.4 cm]{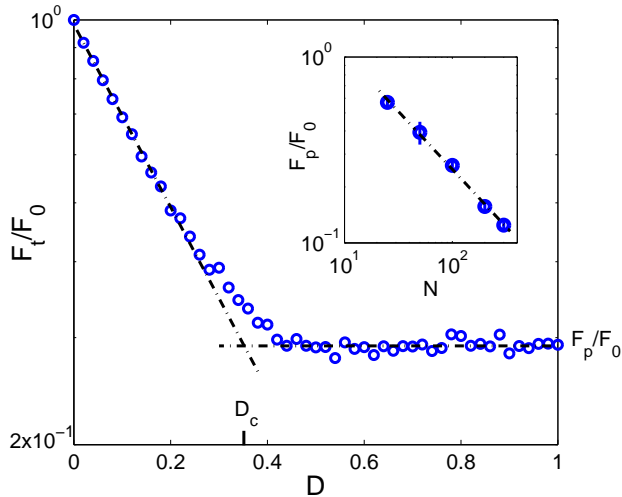}}
\caption{[Color online] Semi-log plot of the transmitted amplitude as a function of $D$ for a long spin chain ($N = 100$ steel:steel dimer cells, with particle sizes chosen to give mass ratio $m_1/m_2 = 0.25$). Straight lines represent fits of the data to Eq.~(\ref{Eq2}). In the inset, we show in logarithmic coordinates the effect of the chain length $N$ (ranging from $N = 25$ to $N = 300$ dimer cells) on the transmitted force amplitude in the high-disorder regime $D \gg D_c$.  We observe a power-law dependence with exponent 0.62 obtained from a fit of the data (in agreement with the theoretically predicted value of $\mu = 3/5$).  
}
\label{Fig2numerics}
\end{figure}

Experiments and simulations (with as many as $N > 500$ cells in the latter) both suggest the existence of two different regimes: At low disorder---i.e., for $D$ smaller than some threshold $D_c$---we see numerically that the transmission of the wave decays exponentially with $D$ (see Fig.~\ref{Fig2numerics}).  As a result, the wave propagation depends on the number of defects in the chain in this regime.  This exponential decay arises in the time-evolution 
of the wave amplitude (as the wave propagates through one defect 
after another). The decay of the wave, in turn, induces background
excitations. Due to the presence of the defects, we observe states that
seem to be somewhat reminiscent of the phenomenon of Anderson localization.  For a 
detailed discussion of acoustic analogs of Anderson localization and excitation of the
pertinent modes in linear disordered acoustic chains, see Ref.~\cite{Maynard}.  We discuss the similarities and differences between the phenomena that we observe and those known from studies of Anderson localization in more detail in Section \ref{Section_discussion}.  Our observation that disorder gradually alters the properties of wave propagation is to be expected, given that this also happens in linear settings. However, for high disorder ($D > D_c$), the response of the chain becomes independent of the level of heterogeneity of the system. This is a fascinating property of highly nonlinear chains; beyond a critical level of disorder $D_c$, the wave behavior is insensitive to the level of heterogeneity in the chain.  As a result, we see (based on our numerical observations) that the peak force of the transmitted wave is well described by
\begin{align}
	F_t &= F_0 e^{-{N D}/{\alpha}} \quad (D \ll D_c)\,, \notag \\
	F_t &= F_0 \frac{\beta}{N^\mu} \quad (D \gg D_c)\,, \label{Eq2}
\end{align}
where $F_t$ is the peak force of the transmitted wave, the original wave's peak force is $F_0 = 1$ by normalization with respect to the maximum force in the perfectly periodic chain (which has $D = 0$), $\mu = 3/5$ is universal, and $\alpha \approx 28$ and $\beta \approx 4.4$ are constants whose values depend on the particle geometries (i.e., their shape) and material properties in the chain.  We measured the values of $\alpha$ and $\beta$ using numerical fitting for our configuration---a large steel:small steel diatomic chain (the mass ratio is $m_1/m_2 = 0.25$).  We show the force transition in both regimes for a large steel:small steel chain with $N = 100$ cells in Fig.~\ref{Fig2numerics}.  (In illustrating our results, we focus the discussion in the remainder of this paper on large steel:small steel chains.  Our results do not depend on this choice, except for particular numerical values when indicated.) In the inset of the figure, we show the dependence of the transmitted force amplitude $F_t$ on the chain length $N$ when $D \gg D_c$.  Namely, $F_t$ decays as a power law in $N$ with a coefficient $\mu$ that reflects the nonlinear relationship between the force and displacement and the 1D nature of the configuration.  The value of the power $\mu$ arises from the energy-force scaling of $E \sim F^{5/3}$ due to the Hertzian contact law (i.e., $F \sim \delta^{3/2}$) and, for example, we would instead obtain $\mu = 1/2$ for a linear contact law (for which $E \sim F^2$).  In the next subsection, we will justify the effect of the chain length when $D \ll D_c$.  In Section \ref{secC}, we discuss the high-disorder regime and the value of $\mu$.


The presence of two regimes with very different dynamics in the aggregate response of the disordered chain raises the question of the origin of the transition between them. Below we analyze the spatio-temporal structures of the propagating waves in both regimes and reveal their physical origins. 



\subsection{Low-Disorder Regime: Solitary Waves with Decaying Energy}

\begin{figure}[tbp]
\centering{
(a) \includegraphics[width= 9.0 cm]{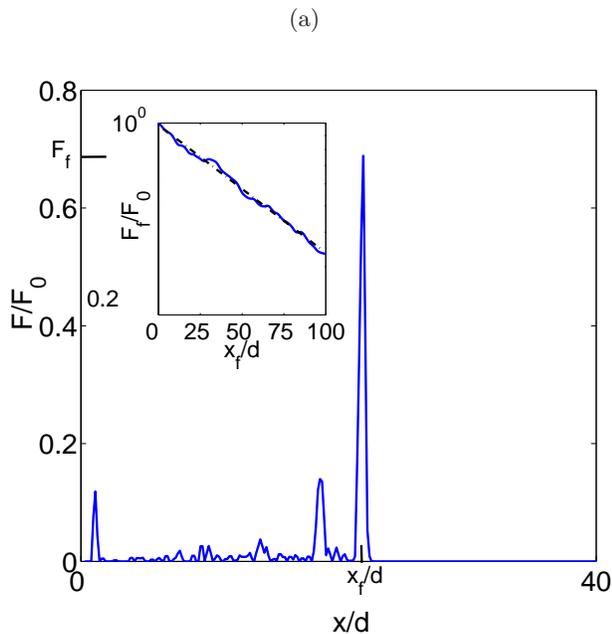}
(b) \includegraphics[width= 8.4 cm]{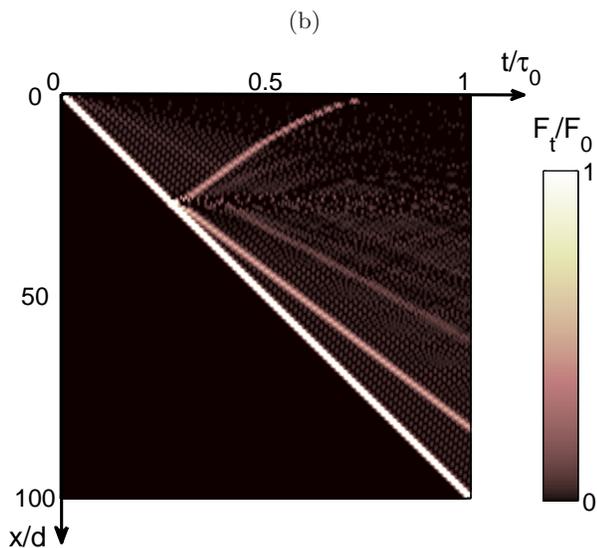}
}
\caption{[Color online] (a) Wave structure in the low-disorder regime ($D < D_c$) for $D = 0.28$ and fixed time. We observe a clean leading pulse with a shape that is similar to the solitary wave that propagates in perfectly ordered chains (which have $D = 0$).  In the inset, we show the exponential decrease [see Eq.~(\ref{Eq4})] of the amplitude of the force as the wave propagates through the chain.  This decrease is induced by the scattering of the solitary wave by the defects in the chain.  As discussed in the text, $x/d$ gives the coordinate normalized relative to the dimer size ($d$ is equal to the sum of the radii of the two particles).  (b) Spatio-temporal depiction of the scattering of the wave due to its propagation past an isolated defect (for a chain with a single defect, so that $D = 0.02$). The quantity $\tau_0 \approx 1.7 \mbox{ ms}$ denotes the time that it takes for the wave to travel through a perfectly ordered chain ($D = 0$).
}
\label{Fig_low}
\end{figure}

For concreteness, we frame our discussion using a steel:steel chain with $m_1/m_2 = 0.25$ and $N = 100$ cells (as in Fig.~\ref{Fig2numerics}), so that $D_c \approx 0.35$.  When $D \ll D_{c}$, the propagating wave consists of a leading pulse that resembles a solitary wave [see Fig.~\ref{Fig_low}(a)]. Indeed, the width and the shape of the leading pulse are similar to those for (fully-ordered) diatomic chains with $D = 0$ \cite{dimer}. However, because of the defects, the amplitude of the solitary wave decays as it propagates through the chain. The inset shows the peak force $F_f$ of the pulse as a function of its position $x_f/d$ in the chain.  This force is averaged over various configurations of the chain with the same level of disorder (i.e., with the same value of $D$). Note that $x_f$ denotes the position of the peak force [so that $F_f = F(x_f)$], and $d = R + r$ (where $R$ is the radius of the first type of particle and $r$ is the radius of the second type of particle) denotes the length of the cell. As the solitary wave propagates, its peak amplitude decays exponentially:
\begin{equation}
	F(x_f) = F_0 e^{-{x_f}/{\xi}}\,. \label{Eq4}
\end{equation}
The length scale $\xi$, which we measure from a fit to the numerical results, is approximately $92 d$ for the large steel:small steel configuration in Fig.~\ref{Fig_low}.  The value of $\xi$ depends on $D$ and on the geometries and material properties of the particles.  We show the exponential decay in the inset of Fig.~\ref{Fig_low}(a).  One might be inclined to 
think of the parameter $\xi$ as an analog of Anderson localization length in disordered linear settings.  A key difference with Anderson localization, however, is that the exponential localization that we observe occurs in time (as the wave propagates through one defect after another), whereas localization of Anderson modes occurs in space.  Keeping this key difference in mind, we nevertheless discuss a possible analogy with linear systems in more detail in Section \ref{Section_discussion}.

The decrease of the wave amplitude is a consequence of the scattering of the wave as it crosses a defect (i.e., a flipped spin) in the chain. To illustrate this process, we show the interaction of the solitary wave with an isolated defect in Fig.~\ref{Fig_low}(b) using a spatio-temporal depiction of the wave as it propagates through a chain with a single defect (so that $D=0.02$).  Observe the two additional waves (in addition to the primary pulse) that are emitted 
at the location of the collision between the incident wave and the defect.

The generation of additional waves in the chain results in a transfer of energy from the leading pulse to the tail \cite{job09}. As proposed in Ref.~\cite{Gredeskul} in the context of weakly nonlinear waves in disordered media, this process can be described by introducing a transmission coefficient $T$ across a single defect.  (The coefficient depends on the geometries and material properties of the particles.)  This allows us to express the energy of the leading pulse after propagation through a defect as
\begin{equation}
	E_{\mbox{after}} = T E_{\mbox{before}}\,.
\label{Eq_T}
\end{equation}
To measure the microscopic parameter $T$, we use diatomic chains with $N=100$ cells 
that contain a single defect, and we measure the amplitude of the wave's force profile $F$ both right before the defect and right after it.  For our calculation, we exploit the fact that the energy in the chain scales with force as $E \sim F^{5/3}$ and that the force scales with the displacement $\delta$ as $F \sim \delta^{3/2}$ for beads with Hertzian contact \cite{nesterenko1}.  The transmission coefficient is then given by $T = [F_{\mbox{after}}/F_{\mbox{before}}]^{5/3}$.  

\begin{figure}[tbp]
\centering{
\includegraphics[width=9.0 cm]{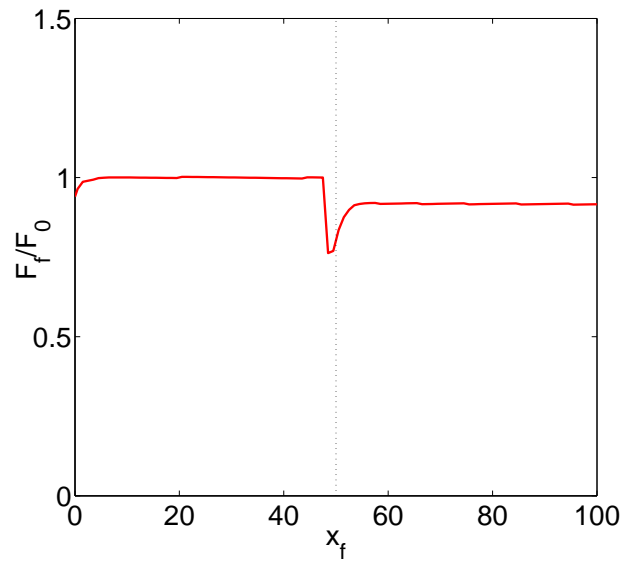}
}
\caption{[Color online] Peak force of the solitary wave as it propagates past an isolated defect. Far from the defect, the wave has the same properties as a solitary wave in the corresponding ordered chain. The presence of the defect only results in a (local) energy loss for the wave as it passes through the defect.  We quantify this energy loss using the transmission coefficient $T = \frac{E_{\mbox{after}}}{E_{\mbox{before}}}=\left(\frac{F_{\mbox{after}}}{F_{\mbox{before}}}\right)^{5/3}$.  
}
\label{Fig_ampl}
\end{figure}

\begin{figure}[tbp]
\centering{
\includegraphics[width=9.0 cm]{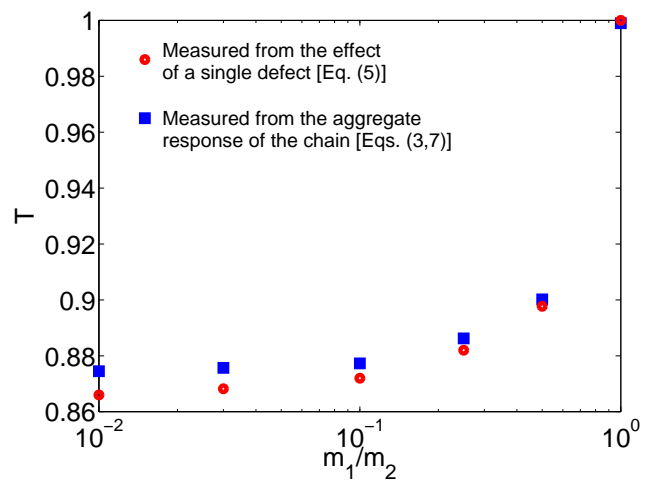}
}
\caption{[Color online] Transmission coefficient 
as a function of mass ratio for the solitary wave that we observe in the low-disorder regime [using Eq.~(\ref{Eq_T})].  The values that we obtained from considering the energy loss when passing through an isolated defect (see Fig.~\ref{Fig_ampl}) show excellent quantitative agreement with those that we obtained from the aggregate transmission properties of the elastic spin chain [using Eqs.~(\ref{Eq2}, \ref{Eq_macro})], which we derived from our model of wave propagation in the low-disorder regime.
}
\label{Fig_T}
\end{figure}

We illustrate this diagnostic in Fig.~\ref{Fig_ampl} for a steel:steel chain with mass ratio $m_1/m_2 = 0.1$. Importantly, the transmission coefficient does not depend on the energy of the incident wave, in contrast to what has been observed in investigations of weakly nonlinear waves \cite{Gredeskul}.  For example, we find that $T \approx 0.882$ for the steel:steel chain of Fig.~\ref{Fig_low} (which has a mass ratio of $m_1/m_2 = 0.25$) and that $T \approx 0.84$ for the steel:Al chain. To illustrate the effect of dimer composition on the value of $T$, we plot $T$ as a function of the mass ratio of the beads in a steel:steel chain.  We indicate these values with red circles in Fig.~\ref{Fig_T}. Note that when $m_1/m_2 = 1$, it is necessarily also true that $T = 1$, as this corresponds to the propagation of a solitary wave in a fully ordered ($D = 0$) diatomic chain.

The aforementioned understanding of the effect of a single defect allows us to consider the full granular chain in the low-disorder regime. Assuming that each defect interacts with the wave independently, we can derive an expression for the amplitude of the transmitted force that results from the successive scattering of the incident wave from each individual defect. For a chain with disorder $D$, the expected number of defects that are encountered by the wave after propagating a distance $x_f$ is given by $\frac{x_f}{2d}D$.  (Note that obtaining this formula requires averaging over multiple chain configurations with the same disorder value.) The pulse energy is reduced to $E = E(t = 0) T^{\frac{x_f}{2 d} D}$.  The relation $E \sim F^{\frac{5}{3}}$ implies that
\begin{equation}
	F(x_f) = F_0 e^{-\frac{x_f}{\xi}}, \qquad \xi = \frac{10 d}{3 D \log(\frac{1}{T})}\,,
\label{Eq_micro}
\end{equation}
where $F_0 = 1$ by normalization. This agrees with the wave structure in Eq.~(\ref{Eq4}) that we obtained numerically. Using $T \approx 0.882$ and $D \approx 0.28$, which was the case for the example in Fig.~\ref {Fig_low}, we obtain $\xi \approx 95 d$ from (\ref{Eq_micro}), in excellent agreement with the fit of the numerical data in the inset in Fig.~\ref{Fig_low}(a).  

We now consider wave propagation along the entire chain by taking $x_f = N d$.  This yields the exponential decay of Eq.~(\ref{Eq2}), with parameter
\begin{equation}
	\alpha = \frac{10}{3 \log({1}/{T})}\,. \label{Eq_macro}
\end{equation}
Equation (\ref{Eq2}) can thereby be used to describe the decrease of the transmitted amplitude 
for $D \ll D_c$.  The value $\alpha \approx 27$ for the steel:steel chain with mass ratio $m_1/m_2 = 0.25$ that we obtained from the transmission coefficient $T \approx 0.882$ using this expression matches the value that we measured directly in numerical simulations. To further test our theoretical results, we compare in Fig.~\ref{Fig_T} the value of $T$ predicted by the exponential decrease of $F_t(D)$ using Eq.~(\ref{Eq2}) with the value that we measured from the interaction of the wave with a single defect.  These results agree quantitatively within $1\%$, confirming that the defects can be treated separately when considering the transmitted force, provided that $D \ll D_c$.


\subsection{High-Disorder Regime: Delocalization of the Waves} \label{secC}

We saw earlier that if the level of disorder in the chain is low (e.g., if $D \ll D_c$), then the disorder $D$ governs the wave transmission through the chain by fixing the number of defects that scatter its energy during propagation. However, when there are too many defects, the amplitude of the transmitted wave becomes robust to the configuration of the elastic spin chain and no longer depends on $D$ (see Fig.~\ref{Fig2}).  A detailed analysis of the wave structure in this high-disorder regime reveals that the spatio-temporal evolution of the wave is now independent of the level of disorder, as the system has reached a level of maximum scattering by the chain heterogeneities and additional defects do not further delocalize the propagating wave.

\begin{figure}[tbp]
\centering{ \includegraphics[width= 8.4 cm]{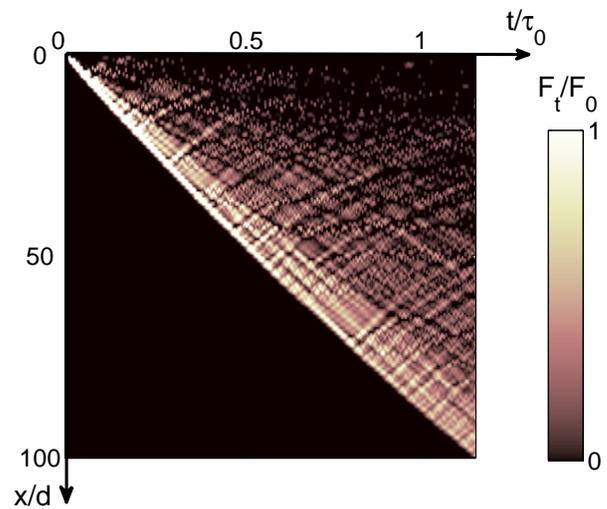} }
\caption{[Color online] Spatio-temporal structure of the wave in the high-disorder regime ($D = 0.70$). As in Fig.~\ref{Fig_low}, the quantity $\tau_0 \approx 1.7 \mbox{ ms}$ denotes the time that it takes for the wave to travel through the perfectly ordered chain.}
\label{Fig_spatio_high}
\end{figure}

\begin{figure}[tbp]
\centering{
\includegraphics[width = 9.0cm]{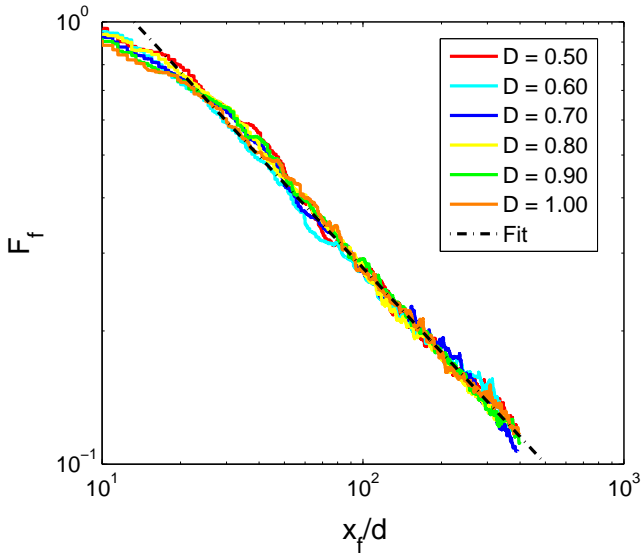}
}
\caption{[Color online] Force amplitude of the leading front of the delocalized wave as a function of its position $x_f$ in the chain.  For each curve, we average over 10 configurations of the chain with disorder level $D > D_c$.  As the wave propagates and after a short transient, the force amplitude decays as a power law with a coefficient of about $0.63$, which is close to the theoretical prediction of $\mu = 3/5$.
}
\label{Fig_high}
\end{figure}

In Fig.~\ref{Fig_high}, we show the basic process of wave delocalization in highly disordered chains.  To obtain each curve in this figure, we fix the disorder level $D > D_c$ and randomly assign the positions of the $N \times D /2$ defects in that chain.  We then measure the force amplitude as the wave propagates through the chain and plot the mean
\begin{equation*}
	\langle F^j(x) \rangle := \frac{1}{10}\sum _{j = 1}^{10} F^j(x)\,,
\end{equation*}
where $F^j(x)$ gives the force amplitude of the $j$th configuration.  As the wave propagates, the force amplitude $F_f = F(x_f)$ of its front decays and the wave broadens in space so that its energy spreads throughout all (continuous) positions $x \in [0, x_f]$.  To quantitatively describe the spatial and temporal dynamics of the wave structure, we average over the different configurations of the elastic spin chain with disorder parameter $D$ in order to determine the primary wave dynamics at a fixed level of heterogeneity.  Because the aggregate transmission of the wave is robust to the level of heterogeneity in the chain, we find that the averaged spatio-temporal wave structure is also independent of $D$.  It is given by
\begin{equation}
	F(x,t) = H(x_{f}-x) F_f e^{-\left(\frac{x_{f}-x}{x_{f}}\right) }\,,
\label{Eq5}
\end{equation}
where $H$ is the Heaviside function, $x_f$ is the position of the front of the delocalized wave, and $F_f = F(x_f)$ is its corresponding amplitude.  As illustrated in Fig.~\ref{Fig_high}, the decay of the force amplitude of the wave front is well described by a power law $F_{f} \sim x_f^{-0.63}$, in excellent agreement with the theoretical prediction of $\mu = 3/5$, as it propagates through the chain.  In contrast to the wave dynamics in the low-disorder regime, the wave scattering in the high-disorder regime is independent of the disorder parameter $D$.

We can capture the main properties of the wave in the high-disorder regime by exploiting the fact that the scattering has reached its maximum level in the chain and approximating the resulting wave structure by a highly delocalized wave in which the elastic energy is spread out evenly among
the dimer cells for which $x \in [0, x_f]$.  (The only cells that are excluded are those that the wave has not yet had enough time to reach.)  In this situation, the scattering in the chain has reached its maximum level because the wave is now fully delocalized.  Additional defects do not cause further delocalization, so the wave properties are robust in this regime to the specific value of $D$.
Such a stationary spatial structure of the wave yields an energy per cell of $\frac{E_{\mbox{tot}}}{x_f/d}$ when the front reaches $x_f$.  Using the relation $F \sim E^{3/5}$ between the peak force and the wave energy yields $F \sim x_f^{-3/5}$ (see the discussion below), which agrees with the numerical observations of Fig.~\ref{Fig_high}. If we consider such stationarity over the full chain---i.e., when the front reaches the end of the chain, so that $x_f = N d$---we obtain the force amplitude of the transmitted wave through the chain.  This gives a plateau whose level is $F \sim N^{-3/5}$ [see Eq.~(\ref{Eq2})], in excellent agreement with the direct numerical simulations shown in the inset in Fig.~\ref{Fig2numerics}.

Although one can observe behavior like that shown in Fig.~\ref{Fig_high} when there is energy equipartition, such a mechanism is not necessary.  Indeed, as one can see from Eq.~(\ref{Eq5}), we do not have equipartition.  Instead, we observe that the wave structure $F(x,t)$ remains stationary while expanding, which is necessary to obtain the features in Fig.~\ref{Fig_high}.  From a mathematical perspective, this implies that we can write the wave structure as 
\begin{equation}
	F(x,t) = F_f(x_f)u\left(\frac{x_f-x}{x_f}\right)\,, \label{scaling}
\end{equation}
where $u$ is the spatial structure of the wave, which expands as the position of the front $x_f$ increases.  The amplitude of the wave and its evolution during the propagation is given by $F_f(x_f)$.  By energy conservation, the integral of the energy density $F(x,t)^{5/3}$ over the portion of chain where the wave is non-zero---i.e., over the interval $[0, x_f]$---must remain constant.  It is thus independent of $x_f$, which implies that the amplitude of the wave decays as $F_f(x_f) \sim x_f^{-3/5}$.  In contrast, in the solitary-wave regime, the structure of the wave is not stationary, and energy is transferred from the leading pulse to the tail of the wave, so that the wave structure cannot be written in the form (\ref{scaling}).  In the high-disorder regime, the fluxes of energy from the leading part of the wave to the tail and those from the tail to the leading part of the wave balance each other, and the wave structure does not change aside from the dilation in Eq.~(\ref{scaling}).


\subsection{Finite-Size Effects}

From our understanding of the low-disorder and high-disorder regimes, we can now analyze the effect of the chain length on the transition from solitary wave propagation (for $D \ll D_c$) to highly delocalized waves ($D \gg D_c$).  We have used the force amplitude $F_t$ of the wave as a function of the level of disorder $D$ in the chain to characterize this transition: The force $F_t$ first decreases exponentially with $D$ and then saturates to a plateau for $D \gg D_c$.  This provides a natural way to define the critical level of disorder $D_c$ as the intersection between both regimes [see Fig.~\ref{Fig2}(a)].  Equating the two expressions in Eqs.~(\ref{Eq2}), which give the force of the incident wave as a function of the chain length in the two regimes, we obtain the critical defect density
\begin{equation}
	D_c = \frac{\alpha}{N}\left[\mu \log(N) - \log(\beta)\right]\,,
\label{Eq_finite}
\end{equation}
where $\alpha$ is expressed in Eq.~(\ref{Eq_macro}) as a function of the transmission coefficient of the wave through one defect and $\mu = 3/5$ is a universal exponent. 
As the chain length increases, the transition occurs for smaller values of $D_c$, and the threshold tends to $0$ as the chain becomes infinitely long.  This result can be interpreted as follows: It is the number of defects that the wave encounters rather than the defect density that determines the type of wave propagation (i.e., whether one observes a solitary pulse or a delocalized one).  As a result, for a very long chain, even a low defect density is able to generate delocalization.  When the force amplitude of the leading pulse is comparable to that of a reflected wave, we are in the delocalized-wave regime.  This occurs when the amplitude of the leading pulse is below a critical level, which happens once it has encountered a given number of defects (recalling that the amplitude decays by a factor of $1 - T$ each time the wave encounters a defect).

\section{Discussion} \label{Section_discussion}

Let us now compare and contrast our observations of the wave properties that we have seen in highly nonlinear chains with those reported previously for linear and weakly nonlinear settings (see Ref.~\cite{Maynard} for a review). In particular, it is interesting to consider the wave-scattering mechanisms that we have observed in elastic spin chains versus the process of Anderson localization that can occur in linear and weakly nonlinear systems \cite{anderson58,andersonstuff}.

We have observed that the heterogeneities present in the granular chain alter 
the wave propagation, so that in the limit of a very long chain, even a low 
defect density is sufficient to prevent wave propagation through the whole system. This intuitive effect is also observed in linear disordered 
chains through the localization in space of the wave close to the excited part of the chain for arbitrary levels of disorder, as ensured by Furstenberg's theorem \cite{Furstenberg}.  (Here, we use the terms ``localization" and 
``localized wave" to contrast these observations with the extended state of the waves that arise in periodic and perfectly ordered linear chains.)  A signature of Anderson localization in linear disordered settings is the exponential 
decay of the wave amplitude as it propagates through the 
chain \cite{Akkermans, Furstenberg}.  This defines the Anderson localization 
length involved in such an evolution law.  As we have reported in this paper, 
elastic spin chains exhibit exponential amplitude 
decay [see Eq.~(\ref{Eq4})] for low levels of disorder ($D \ll D_c$). 
It is then perhaps tempting to consider
the localization length $\xi$ defined in Eq.~(\ref{Eq_micro}) 
as a nonlinear counterpart of the Anderson localization length for highly 
nonlinear settings. However, the reader is cautioned towards such a connection, as the decay of the amplitude is connected to the wave propagation
over time (as additional defects are encountered) and does not really
relate to the structure of the spatial profile of the wave.
 Furthermore, recall (see, for example, the discussion of
Section 4.2 in Ref.~\cite{lepri}) that FPU-type systems do not admit a
genuine form of Anderson localization (at least for small frequencies) unless a harmonic on-site potential is added to each node in the lattice. 
As another point of comparison, we have provided in this paper an expression for $\xi$ as a 
function of the defect density $D$ and the scattering characteristics of one 
isolated defect $T$, which is also possible for linear disordered chains in 
some specific situations \cite{Akkermans,Lambert}.

Localization in highly nonlinear systems can be fundamentally different than localization in linear and weakly nonlinear systems.  In particular, we observe in granular chains a fascinating property (which is a central result of our study):  Beyond a critical level of disorder, the localization length saturates, so that the wave dynamics remains insensitive to the level of heterogeneity in the chain in this regime. In addition, the decay of the wave amplitude does not follow an exponential relation but instead follows a power-law relation (see Fig.~\ref{Fig_high}), with a power fixed by the type of nonlinear interaction between nearest neighbors in the chain. This remarkable and unexpected effect arises from the balance between the flux of energy from the leading pulse to the tail of the wave and the flux in the opposite direction. It results in a stationary wave structure that, combined with an argument based on energy conservation in the chain (see Section \ref{secC}), suggests the possibility of power-law decay (in time) of the wave amplitude for sufficiently high levels of disorder. How and when this condition of flux balance is achieved still remains to be established. However, our investigation of the wave behavior in both the low- and high-disorder regimes gives an expression of the critical level $D_c$ of disorder corresponding to the transition from exponential to power-law decay of wave amplitudes [see Eq.~(\ref{Eq_finite})]. It is noteworthy that for sufficiently long chains, the system experiences the second type of decay (because $D_c \sim \log(N)/N$).  Consequently, this regime is expected to dominate for long times and large system sizes.  



\section{Conclusions}  \label{sec7}

We have introduced the concept of elastic spin chains, which we constructed using randomly-oriented arrangements of dimer cells in one-dimensional granular crystals. We quantified the level $D$ of disorder in the chain by counting the number of spins with the same orientation.  In order to characterize the wave propagation properties, we calculated ensemble averages of relevant quantities over different chain configurations with the same value of $D$---i.e., configurations with the same number of up and down spins. This allowed us to quantitatively characterize the effect of disorder on the propagation of highly nonlinear pulses. For low levels of disorder, we observed the propagation of a solitary wave with exponentially decaying amplitude.  
We observed amplitude decay in time (as the wave propagates through one defect 
after another).  The ``localization length'' characterizing the exponential 
amplitude decay in the low-disorder nonlinear setting can be expressed as a 
function of the level of disorder $D$ and the scattering properties of an 
isolated defect (i.e., when one dimer has a different spin from all of the 
other dimers).  As a result, the transmission coefficient through the entire 
chain decays exponentially with the level of disorder $D$. 
Beyond a critical level of disorder $D_c$, this transmission coefficient 
becomes independent of $D$ and saturates at a constant value. In this regime, 
in which the wave behavior remains insensitive to the level of disorder, the 
wave's spatial structure is much more extended.  Moreover, the propagation of 
the wave is no longer governed by the propagation of an isolated solitary 
wave, as it was the low disorder regime, but it instead resembles a train of 
smaller waves. In addition, the decay of the wave amplitude follows a power 
law instead of an exponential law, with the power fixed by the type of 
nonlinear interaction between nearest neighbors in the chain. This 
effect appears to
arise from a balance between the flux of energy going from the 
leading pulse to the tail of the wave and the flux going in the opposite 
direction. How and when this condition of flux balance is achieved remains 
to be established, and additional studies of localization phenomena in highly 
nonlinear settings should lead to interesting insights into the dynamics of 
disordered nonlinear systems.



\vspace{.1 in}

\section*{Acknowledgements}

We thank D. Allwright, D.~K. Campbell, E.~J. Hinch, E. L\'{o}pez, and G. Refael for useful discussions and K. Whittaker for help with experiments.  We also thank the European Union's (PhyCracks project) Marie Curie outgoing fellowship (LP), NSF-CMMI (CD), NSF-CAREER (CD, PGK), NSF-DMS (PGK), the Jardine Foundation (YML), and Exeter College in Oxford (YML) for support.


\end{document}